\documentclass[structabstract]{aa}

\usepackage{txfonts}
\usepackage{graphicx}
\usepackage{aalongtable}
\usepackage{multirow}
\usepackage{natbib}
\bibpunct[; ]{(}{)}{,}{a}{}{;} 

\begin{document}

\title{Near-Infrared Evidence for a Sudden Temperature Increase in Eta Carinae
}

\author{Andrea Mehner\inst{1}
\and Kazunori Ishibashi\inst{2} 
\and Patricia Whitelock\inst{3,4}
\and Takahiro Nagayama\inst{2}
\and Michael Feast\inst{3,4}
\and Francois van Wyk\inst{3}
\and Willem-Jan de Wit\inst{1}
    }

\offprints{A. Mehner, \email{amehner@eso.org}}

\institute{ESO, Alonso de Cordova 3107, Vitacura, Santiago de Chile, Chile 
\and Division of Elementary Particle Physics and Astrophysics, Graduate School of Science, Nagoya University, Nagoya, 464-8602, Japan
\and South African Astronomical Observatory, PO Box 9, Observatory, South Africa
\and Astronomy, Cosmology and Gravity Centre, Astronomy Department, University of Cape Town, 7701, Rondebosch, South Africa
}


\abstract {} {Eta Car's ultra-violet, optical, and X-ray light curves and its spectrum suggest a physical change in its stellar wind over the last decade. It was proposed that the mass-loss rate decreased by a factor of about 2 in the last 15 years. We complement these recent results by investigating the past evolution and the current state of $\eta$ Car in the near-infrared (IR).} {We present $JHKL$ photometry of $\eta$ Car obtained at SAAO Sutherland from 2004--2013 with the Mk~II photometer at the 0.75-m telescope and $JHK_{\textnormal{\scriptsize{s}}}$ photometry with SIRIUS at the 1.4-m IRSF telescope from 2012--2013. The near-IR light curves since 1972 are analyzed.} {The long-term brightening trends in $\eta$ Car's $JHKL$ light curves were discontinuous around the 1998 periastron passage. After 1998, the star shows excess emission above the extrapolated trend from earlier dates, foremost in $J$ and $H$, and the blueward, cyclical progression in its near-IR colors is accelerated. The near-IR color evolution is strongly correlated with the periastron passages. After correcting for the secular trend we find that the color evolution matches an apparent increase in blackbody temperature of an optically thick near-IR emitting plasma component from about $3500$ to $6000$~K over the last 20 years.} {
We suggest that the changing near-IR emission may be caused by variability in optically thick bremsstrahlung emission. Periastron passages play a key role in the observed excess near-IR emission after 1998 and the long-term color evolution. We thus propose as a hypothesis that angular momentum transfer (via tidal acceleration) during periastron passages leads to sudden changes in $\eta$ Car's atmosphere resulting in a long-term decrease in the mass-loss rate.} 


\maketitle

\section{Introduction}
\label{introduction}

Eta Car is, with $M > 100~M_\odot$ and $L \sim 5\times10^6~L_\odot$, one of the most massive and most luminous stars in our Galaxy. During its {\it Great Eruption} almost two centuries ago, its luminous energy output rivaled that of a supernova (SN) as it expelled more than $10~M_\odot$ \citep{2003AJ....125.1458S}, which are now observed as the Homunculus nebula. Because of its proximity to us, $\eta$ Car is one of a few ``SN impostors'' whose recovery and continuing instability can be analyzed in detail. Diverse datasets cover the last four centuries, but despite numerous multi-wavelength studies there is still no consensus on the nature of $\eta$ Car and the cause of its Great Eruption (\citealt{1997ARA&A..35....1D,1999ASPC..179.....M,2005ASPC..332.....H,2012ASSL..384.....D}, and references therein). 
It is now generally believed that $\eta$ Car is a binary system (\citealt{1996ApJ...460L..49D,2000ApJ...528L.101D}; cf.\ \citealt{1997NewA....2..387D,2000ApJ...530L.107D}). 
However, the details of the orbital parameters and the companion's influence on the observables are still debated. For recent discussions, see \citet{2012MNRAS.423.1623G,2012MNRAS.420.2064M,2008MNRAS.388L..39O}, and for an alternative model \citet{2008MNRAS.390.1751K}.

Most of the flux from the $\eta$ Car system emerges at mid- to far-infrared (IR) wavelengths and is reprocessed short-wavelength radiation from the central stars by cool dust in the Homunculus nebula (e.g., \citealt{2003AJ....125.1458S}, and references therein).
At optical wavelengths we observe the light originating from $\eta$ Car's stellar wind, the Weigelt knots and similar close ejecta, and dust-scattered light from the Homunculus nebula. 
The angular size of the system is smallest at near-IR wavelengths and near-IR photometry is therefore particularly valuable for ground-based monitoring of the central source. Images reach a minimum size around $2~\mu$m \citep{1973ApL....13...89G,1989MNRAS.241..195A} but cyclical variations in the spatial extent exist. \citet{2000ApJ...529L..99S} found that the central core changed its morphology in $K$ band between 1995 and 1998, appearing more pointlike close to periastron in 1998.
In this paper we discuss the near-IR light curves obtained from large aperture photometry of the entire $\eta$ Car system. The near-IR radiation arises from more than one component, but is likely dominated by free-free emission from the stellar wind \citep{2004MNRAS.352..447W,2005ASPC..332..115W}. Other sources include scattered light from the Homunculus and from close ejecta like the Weigelt knots \citep{1986A&A...163L...5W,1995AJ....109.1784D,2012ASSL..384..129W}, emission lines ($<$10\%, mostly \ion{H}{I} and \ion{He}{I}), and at $K$ and $L$ band hot circumstellar dust ($T_{\textnormal{\scriptsize{dust}}} \gg 200$~K) contributes. Conceivably all components play a role but the importance of each source is still debated. The interpretation of data on $\eta$ Car is difficult because different types of observation refer to physically separate and distinct regions and may experience different extinctions \citep{1995AJ....109.1784D}.

The near-IR light curves show variations on different time scales: secular brightening and color evolution, quasi-periodic variations associated with the 5.5~yr orbit of the companion star, and variations on time scales of a few weeks to months.
Observed variations are likely a combination of several processes, such as the decreasing extinction with the thinning of the expanding Homunculus nebula, dispersion and possibly accretion of circumstellar and wind material by the companion star, varying dust production and destruction, and changes in the stellar wind (e.g., altered stellar rotation and mass-loss rate, a wind cavity around the companion, and altered photoionization of the primary wind due the companion).  

The gradual brightening of $\eta$ Car in the visual and near-IR since the 1940s is generally attributed to a gradual decrease in extinction as the dust formed in the Great Eruption disperses (e.g., \citealt{1971MNRAS.154..415D,1994A&A...283...89V}). The rate of brightening was about the same in the visual and near-IR and the dust grains were therefore thought to be predominantly large producing near-neutral extinction. However, the accelerated rate of brightening after 1997, especially in the ultra-violet (UV), is incompatible with a simple model of decreasing extinction as a result of the steady expansion of a dust shell \citep{1999AJ....118.1777D,1999A&A...346L..33S,2004A&A...423L...1V,2005AJ....129..900D}. \citet{2004MNRAS.352..447W} noted that the near-IR brightening was faster and that the color changes have been greater for the two cycles in 1992.5--2003 than for the two previous ones. 
The simplest interpretation of $\eta$ Car's recent brightening at UV to near-IR wavelengths, its altered X-ray light curve, and the major spectral changes observed in the stellar wind features is a decrease in its mass-loss rate \citep{2009ApJ...701L..59K,2010ApJ...725.1528C,2010ApJ...717L..22M,2012ApJ...751...73M}. However, conceivably these observations could be accounted for by line-of-sight effects  \citep{2012MNRAS.423.1623G,2012ApJ...759L...2G,2013arXiv1310.0487M}. A scenario can be envisioned in which we previously looked down the wall of the wind-wind cavity with a large column density. A small change in the stellar parameters could have changed the wind cone opening angle, resulting in lower column density along our line-of-sight.

In this paper we investigate $\eta$ Car's current state in the near-IR to complement the recent work in other wavelength regions.
Eta Car has been monitored in $JHKL$ at the South African Astronomical Observatory (SAAO) at Sutherland for more than 40 years starting in 1972 \citep{1983MNRAS.203..385W,1994MNRAS.270..364W,2001MNRAS.322..741F,2004MNRAS.352..447W}. We present the most recent $JHKL$ photometry obtained with the Mk~II photometer at the 0.75-m telescope between 2004--2013. The Mk~II photometry is the most consistent dataset of $\eta$ Car, spanning a time-baseline of several decades and is thus uniquely suited to analyze secular changes of this stellar  system.  However, 
there are plans to close down the 0.75-m telescope and a temporal overlap of observations with another system is essential to assure the continuation of these legacy observations. 
Thus, in 2012, we started a $JHK_{\textnormal{\scriptsize{s}}}$ monitoring program with the Simultaneous-3color InfraRed Imager for Unbiased Survey (SIRIUS) at the InfraRed Survey Facility (IRSF) at SAAO. 

In this paper we stress the importance of the binary nature of $\eta$ Car on the observables. The long-term evolution of $\eta$ Car's light curve and colors may be foremost a result of changes in the stellar system (either the stars themselves or the surrounding material) triggered at periastron passages.  \citet{2004MNRAS.352..447W} already noted that the 1998 periastron was a phase of rapid change and that the near-IR colors change discontinuously at periastron passages and thus may be driven by some physical process associated with the orbital period. 
In Section \ref{obs} we describe the observations and data analysis. In Section \ref{results} we present our results, followed by a discussion, Section \ref{discussion}, and the conclusion, Section \ref{conclusion}.

\section{Observations and data analysis}
\label{obs}

 \begin{figure*}
\resizebox{\hsize}{!}{\includegraphics{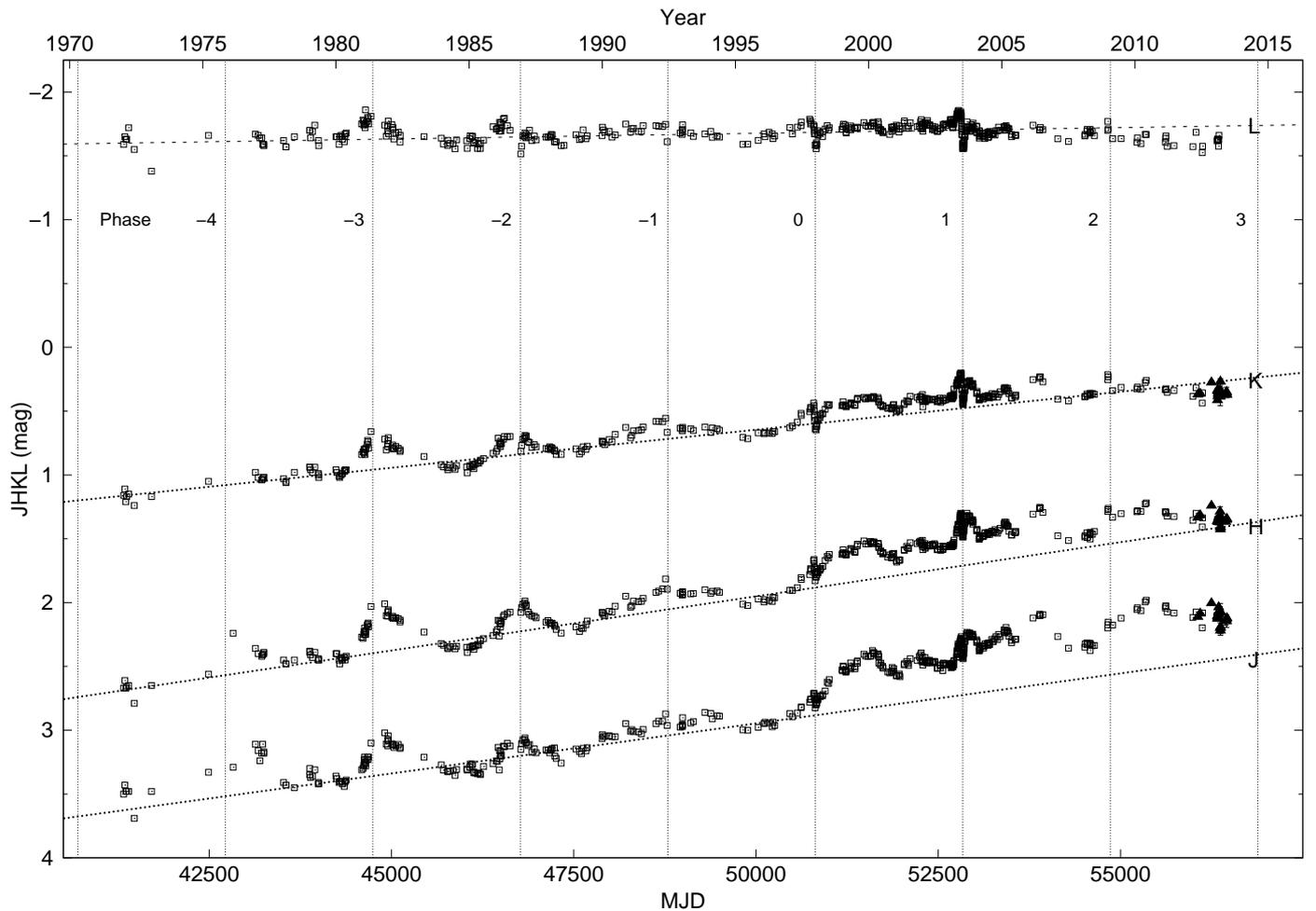}}
     \caption{$JHKL$ light curves from 1972--2013 (open squares: Mk~II, filled triangles: SIRIUS).  The errors at $JHK$ are less than $\pm 0.03$~mag, while those at $L$ are about $\pm 0.05$~mag. Dotted lines show linear regression trends based on the pre-1998 data. Vertical dotted lines indicate periastron passages (the corresponding phases are indicated). In $J$, $H$, and to a lesser degree in $K$ and $L$, $\eta$ Car did not follow the long-term linear brightening trends after the 1998 periastron passage.}
     \label{hist_light curve}
\end{figure*}

\subsection{$JHKL$ photometry with the Mk~II photometer}

A near-IR monitoring campaign of $\eta$ Car has been in progress at SAAO since 1972 \citep{1983MNRAS.203..385W,1994MNRAS.270..364W,2001MNRAS.322..741F,2004MNRAS.352..447W}. Observations were carried out regularly with the Mk~II photometer using a 36\arcsec\ aperture encompassing the entire Homunculus nebula, which has an extension of about 20\arcsec.
Observations from 1972 to 1973 were conducted at the 18-in, the 1-m, and the 1.9-m telescopes. Subsequent observations were carried out at the 0.75-m telescope (the detector was changed only once in 1978 April). The very long time-baseline of more than 40 years of consistent observations makes this dataset invaluable. 

The Mk~II photometry from the years 1972--2004 was published in \citet{1983MNRAS.203..385W,1994MNRAS.270..364W,2001MNRAS.322..741F}, and \citet{2004MNRAS.352..447W}. The SAAO IR standard system \citep{1990MNRAS.242....1C} was used and uncertainties range from of about 0.03~mag in $JHK$ to 0.05~mag in $L$. Here we present the most recent Mk~II photometry from 2004--2013 (Table \ref{table:phot}).  For details about the observations and data reduction, see the papers cited above.

\subsection{$JHK_{\textnormal{\scriptsize{s}}}$ photometry with SIRIUS}

In 2012 May we started a near-IR monitoring campaign of $\eta$ Car with the SIRIUS camera at the 1.4-m IRSF telescope at SAAO \citep{2000MNSSA..59..110G}.\footnote{IRSF is a joint project of Nagoya University, Kyoto University, the National
Astronomical Observatory of Japan (NAOJ), and SAAO.} SIRIUS is a simultaneous 3-channel $JHK_{\textnormal{\scriptsize{s}}}$ camera with a field of view of 7.7\arcmin$\times$7.7\arcmin\ and a pixel scale of 0.45\arcsec/pixel \citep{1999sf99.proc..397N,2003SPIE.4841..459N}.
The data include several hours of consecutive observations to explore the potential presence of hourly to daily variations, which we have not identified. Note, however, that we are only sensitive in our data analysis to variation larger than $\sim$0.01~mag. Simultaneous observations with the Mk~II photometer were performed to determine the transformation between the two photometric systems.

Each observation consisted of 10 dithered exposures to correct for bad pixels.
A nearby field with few stars was observed within 30 minutes from the science exposures to create a sky image.
Observations had to be carried out with neutral density filters to avoid overexposure. This made it impossible to determine $\eta$ Car's magnitudes based on relative photometry to other stars in the field of view. Instead, we determined the extinction coefficients and zero points from standard stars that were observed each night with airmasses ranging from 1.0 to 2.4. 
To facilitate the comparison with the Mk~II photometry we chose two standards (BS4450 and BS4382) from the same list of standards  \citep{1990MNRAS.242....1C} and with similar near-IR magnitudes as $\eta$ Car. None of the standard stars has $\eta$ Car's color.

The data were reduced using the external IRAF package SIRIUS (v.09).
Master dark frames were created and subtracted from the science frames. 
For the flat field response we created differential images of twilight frames without neutral density filters (to reach reasonable count levels), which were median-normalized and median-combined. Eta Car and the standard stars were always centered on the same pixel areas and any vignetting or pattern introduced by the neutral density filters are thus not significant for the resulting photometry.
We mitigated the effect of an inherent vertical discontinuity in the bias count levels across the array boundaries to the flat field response image by rejecting some differential images that inherited the central vertical step.
Sky frames were subtracted from each science frame and the dithered exposures were re-centered and combined. 

The photometry was carried out with standard IRAF tools.  We chose an aperture of 36\arcsec\ consistent with the analysis performed by \citet{1983MNRAS.203..385W,1994MNRAS.270..364W,2001MNRAS.322..741F}, and \citet{2004MNRAS.352..447W}. 
To compare the SIRIUS and the Mk~II photometry we explored two different methods. First, we determined $\eta$  Car's $JHK_{\textnormal{\scriptsize{s}}}$ magnitudes in the SIRIUS system and converted them to the Carter system. There is no published direct transformation formula between the SIRIUS and Carter photometric systems and we thus performed a series of transformations.
Second, in 2013 February we conducted simultaneous observations with SIRIUS and Mk~II and derived a linear transformation for $\eta$ Car between the two systems ($J_{\textnormal{\scriptsize{Mk~II}}} \approx J_{\textnormal{\scriptsize{SIRIUS}}} + 0.101$~mag, 
$H_{\textnormal{\scriptsize{Mk~II}}} \approx H_{\textnormal{\scriptsize{SIRIUS}}} - 0.047$~mag, 
$K_{\textnormal{\scriptsize{Mk~II}}} \approx K_{\textnormal{\scriptsize{SIRIUS}}} - 0.076$~mag). 
The first method resulted in very good results for $J$ and $H$ but we systematically underestimated the resulting $K$ magnitude by 0.04~mag. We thus adopted the linear transformation obtained from the simultaneous observations. Note, however, that $\eta$ Car's color is changing (secular and cyclical changes) and that this transformation will need to be revised in the future.

\section{Results}
\label{results}

\subsection{The JHKL light curves 1972--2013}

Eta Car has been monitored for more than 40 years in $JHKL$ at SAAO. 
Until 1998 the light curves showed an approximately linear brightening trend, superimposed with quasi-periodic variations, see Figure \ref{hist_light curve}.  The overlying quasi-periodic variations associated with the 5.5~yr orbit of the companion star render it difficult to determine the rate and confirm the cause of this secular brightening. 
We determined linear regression lines by selecting only data points obtained at the 0.75-m telescope between 1973 and 1998 at phases 0.50--0.85 in the 5.5~yr cycle.\footnote{When referring to ``phase'' in the 5.5~yr cycle, we use a period of 2023.0 days, see the appendix in \citet{2011ApJ...740...80M}. We denote time close to periastron passage by $t$, such that $t = 0$ at MJD 54860.0 (2009 January 29), MJD 52837.0 (2003 July 17), etc. Periastron most likely occurs within the range $t \approx -15$ to $+15$ days.} 
This phase interval was chosen because of the minimal influence of the companion star during the epoch around apastron. Also, the near-IR excess observed at periastron subsides around phase 0.5.
We find that the star brightened by $\Delta J = 0.029\pm0.001$~mag~yr$^{-1}$, $\Delta H = 0.031\pm0.001$~mag~yr$^{-1}$, $\Delta K = 0.022\pm0.001$~mag~yr$^{-1}$, and $\Delta L = 0.003\pm0.001$~mag~yr$^{-1}$.
The linear regression lines are shown in Figure \ref{hist_light curve} to visualize this long-term brightening trend.

A strong discontinuity in the long-term near-IR brightening trend is observed around the 1998 periastron passage. The system suddenly showed accelerated brightening, most pronounced in $J$ and $H$. For the last 15 years, the light curve in $J$ shows an average excess emission of about 0.35~mag above the extrapolation from the previous trend. The light curve in $H$ had recovered by 2013 to the brightness expected from the long-term trend. In $K$ and $L$, $\eta$ Car showed only a small or no amount of excess emission and in 2013 the star has become fainter in both wavebands than expected from the previous trends. 
The light curves alone do not reveal whether the secular brightness increase and the break in 1998 are related to processes occurring at periastron. Yet, it is suggestive that the 1998 discontinuity coincides with a periastron passage.

\subsection{Long-term and cyclical color variations}

 \begin{figure}
\centering
\resizebox{\hsize}{!}{\includegraphics{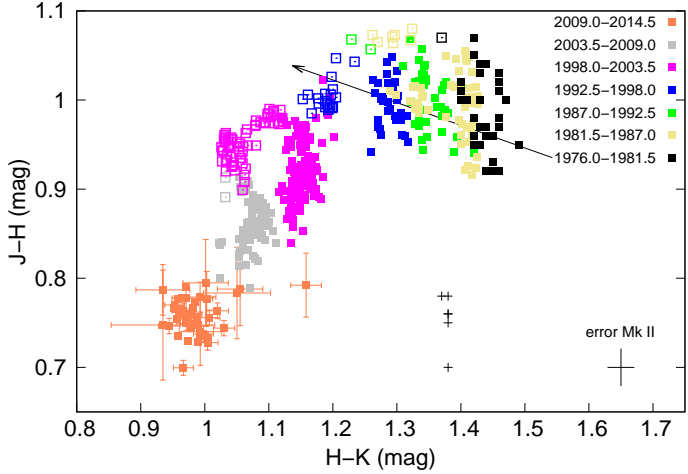}}
     \caption{$J$--$H$ versus $H$--$K$ color-color diagram from 1976--2013. The key refers to the time windows of orbital cycles and not to the actual observation dates. Each cycle is displayed with a different color code to visualize the long-term color evolution and the impact of the periastron passages. A cycle here consists of phases 0.05--1.05, 1.05--2.05, etc.\ to include only one periastron passage per cycle. Open squares of the same color show data points close to periastron ($t = -40$ days to $t = +100$ days).
The arrow indicates the expected evolution from 1970--2014 based on the pre-1998 data. The data points represented by the black plus symbols are from observations obtained before 1978 with a different detector.}
     \label{color-color}
\end{figure}

Eta Car's near-IR color evolution reveals a strong discontinuity around the 1998 periastron passage.
Figure \ref{color-color} shows a color-color diagram ($J$--$H$ versus $H$--$K$) from 1976--2013, covering seven orbital periods. Earlier observations are not included in the following discussion for consistency reasons (they were obtained at different telescopes). The outlying data points of cycle 1976--1981.5 (the black plus symbols in Figure \ref{color-color}) are from observations obtained before 1978 with the previous detector (a lead sulphide detector). The change to the indium antimonide detector still in use did not alter significantly the  $HKL$ magnitudes obtained but $J$ became fainter by $\sim$0.20~mag \citep{1983MNRAS.203..385W}. A possible reason is that the sensitivity of the old detector may have been greater at the short wavelength end of the $J$ band and thus may have been more sensitive to the \ion{He}{I} 10830~\AA\ emission line. The data points are presented here to show how relevant the question of detectors/filters is when combining different data sets.

 \begin{figure}
\centering
\resizebox{\hsize}{!}{\includegraphics{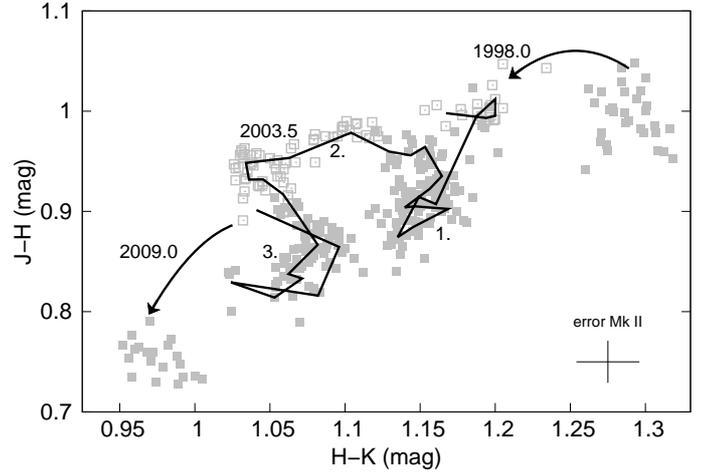}}
     \caption{$J$--$H$ versus $H$--$K$ color-color diagram. Filled squares are observations outside periastron passages (`1.': phases 0.05--0.98, `3.': phases 1.05--1.98). Open squares are observations during periastron (`2.': phases 0.98--1.05, i.e., 2003.5 periastron). The solid curve highlights the quasi-cyclical cycloidal pattern.}
     \label{color-variations}
\end{figure}

Between 1976 and 1998, $J$--$H$ was almost constant with  $\Delta (J$--$H) \sim 0.002$~mag~yr${^{-1}}$. After 1998,  $J$--$H$ is evolving to the blue at a rate of $\Delta (J$--$H) \sim -0.013$~mag~yr${^{-1}}$. This trend appears to be primarily induced at the periastron passages starting in 1998. $J$--$H$ varies by about 0.1~mag throughout the 5.5~yr cycle but shows a sudden increment on the order of 0.1~mag towards the blue during periastron. $H$--$K$ had been slowly evolving towards the blue at a rate of $\Delta (H$--$K) \sim -0.009$~mag~yr${^{-1}}$ since 1976.  After 1998, $H$--$K$ is evolving towards the blue at a faster rate of $\Delta (H$--$K) \sim -0.016$~mag~yr${^{-1}}$. 
As in the case of $J$--$H$, the blueward evolution of $H$--$K$ appears to be mostly triggered at periastron passages; a fast (on the time scale of a few weeks) change to the blue at each periastron is evident in Figure \ref{color-color} (see also Figures 3 and 4 from \citealt{2004MNRAS.352..447W}). 

The long-term color evolution (filled squares in Figure \ref{color-color}) is interspersed by a cyclical pattern. A cycloidal, counter-clockwise 5.5~yr pattern is outlined in the color-color diagram, see Figure \ref{color-variations}. Shortly after periastron, $J$--$H$ and $H$--$K$ vary gradually over the next few years (step 1.\ in  Figure \ref{color-variations}). $J$--$H$ decreases to a minimum value before becoming redder again. The movement to the red is not in the middle of the cycle, but about 1~yr before the next periastron. During this epoch, $H$--$K$ scatters within about 0.05~mag but shows little directed evolution. At the end of a cycle,  $J$--$H$ reaches again about its reddest value of that cycle. Then, within 3--4 weeks, around periastron, $H$--$K$ evolves about 0.1~mag to the blue at almost constant $J$--$H$ (step 2.). The next cycle starts and $J$--$H$ becomes bluer  (step 3., which corresponds to the same stage in the cycle as step 1.). For orbital cycles prior to 1992--1998 this pattern traces a horizontal cycloidal structure in the color-color diagram ($\Delta (J$--$H) \approx$ constant), while for cycles afterwards there is also a vertical trend  ($\Delta (J$--$H ) < 0$).

 \begin{figure}
\resizebox{\hsize}{!}{\includegraphics{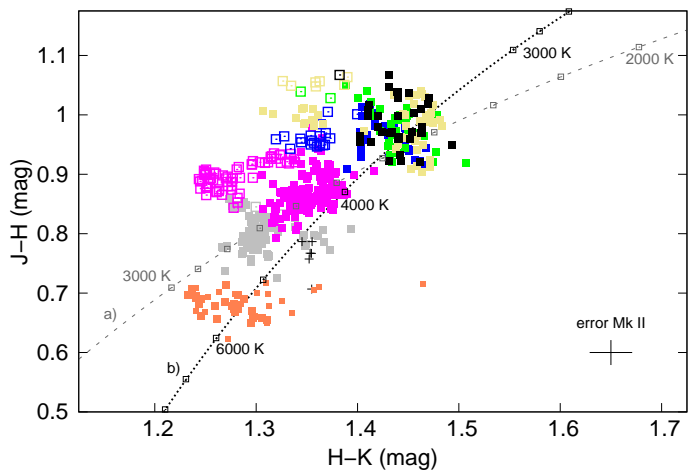}}
     \caption{$J$--$H$ versus $H$--$K$ color-color diagram as in Figure \ref{color-color} (the same symbols are used) but corrected for the secular brightening. The dashed gray curve, a), traces the expected colors of a blackbody reddened by $E(J$--$H)=0$~mag and  $E(H$--$K)=0.863$~mag, the dotted black curve, b), shows a blackbody reddened by $E(J$--$H)=0.4$~mag and  $E(H$--$K)=1.2$~mag, see text. The temperature steps, indicated with open squares along the blackbody functions, are 100~K between 2000--3000~K and 1000~K for higher temperatures.}
     \label{cycle-trendcorr}
\end{figure}

Under the assumption that the long-term linear brightening trend before 1998 is due to the dissipation of dust as the Homunculus nebula expands, we correct the light curves accordingly to obtain the near-IR excess emission. For simplicity, we do not consider a potential non-linear brightening trend.
Figure \ref{cycle-trendcorr} shows the same color-color diagram as Figure \ref{color-color} but corrected for the long-term (linear) trend. The filled squares show the data points outside periastron passages and represent $\eta$ Car's long-term color evolution. The open squares show the data points close to periastron, which lie in a separate region in color-color space. 

Until 1998, the data points fall within the same part of color-color space, but starting around the 1998 periastron passage a long-term blueward color evolution sets in. 
In the following we ignore the cyclical variations (especially step 2.\ in Figure \ref{color-variations}). They are likely caused by photoionization of the primary wind by UV radiation of the companion star (or possibly the primary star itself) and/or a wind cavity around it. Instead we focus on the long-term blueward color evolution, which resembles a change in blackbody temperature. This suggests that we observe a change in a (significant) contribution from an optically thick near-IR emitting plasma radiating nearly at blackbody temperature. Blackbody temperature functions for two different sets of extinction values are displayed in Figure \ref{cycle-trendcorr}. The extinction towards $\eta$ Car is uncertain, but because of the similar long-term trends in $J$ and $H$ we assume a minimum value for $E(J$--$H)$ of 0~mag. A reddening of $E(J$--$H)  \approx 0$~mag and $E(H$--$K)=0.863\pm0.003$~mag then fits $\eta$ Car's data points (curve {\it a} in Figure \ref{cycle-trendcorr}). In this case the blueward trend of $\eta$ Car from 1998 to 2013 resembles an apparent blackbody temperature change from 2400~K to 3000~K. Note that this is a lower limit  in the case that no or a negligible amount
	of optically thin plasma is involved in the emission process.
In addition, we use a simple regression test to search for the set of extinction values that best describes the blueward trend (including only phases 0.5--0.85 to minimize the influence of step 2.\ in Figure \ref{color-variations}). We
find a reddening of $E(J$--$H)=0.4\pm0.1$~mag and $E(H$--$K)=1.2\pm0.1$~mag.\footnote{The quoted errors are systematic errors reflecting the spread of the selected data points around the blueward trend.}
In this case, $\eta$ Car's blueward trend resembles an apparent blackbody temperature change from 3500~K to 6000~K (curve {\it b} in Figure  \ref{cycle-trendcorr}).

As already noted by \citet{2004MNRAS.352..447W}, periastron passages appear to play a key role in the color evolution of $\eta$ Car.
The long-term blueward evolution in $H$--$K$ and, after 1998, also in $J$--$H$ occurs in discontinuous steps close to periastron. The 1998 periastron passage may have been particularly important leading to a significant change in the system that can be observed in the near-IR light curves and color evolution.  
The 5.5-yr quasi-periodic variations make it difficult to determine whether the apparent secular color trends are constant or occur in discontinuous episodes close to periastron. 
However, the involved time scales of only a few weeks support the idea that during periastron permanent changes are induced in the system colors, i.e., the system does not recover to its ``out-of-periastron'' state. 
This observational evidence suggest that the system is changed by the periastron events.

\subsection{Cycle-to-cycle variations}

 \begin{figure}
\resizebox{\hsize}{!}{\includegraphics{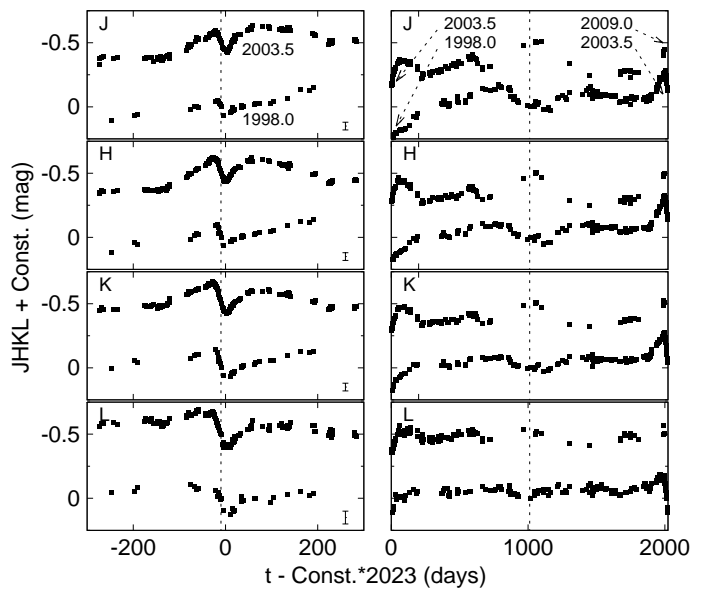}}
     \caption{$JHKL$ light curves showing the cycle-to-cycle differences of the two cycles between 1998.0 and 2009.0. The light curves were not corrected for the linear brightening trend. The vertical lines indicate the mid-decline in the light curve during the 1998 periastron passage (left panels, typical error bars are displayed in the lower right corner) and apastron (right panel).}
     \label{cycle-to-cycle}
\end{figure}

The $JHKL$ light curves show quasi-periodic variations of 5.5~yrs associated with the orbit of the companion star. Figure \ref{cycle-to-cycle} compares the light curves during the periastron passages in 1998.0 and 2003.5 and the mid-cycle light curves from the two cycles between 1998.0 and 2009.0. Variations occur simultaneously in all 4 bands but significant cycle-to-cycle differences exist. 

Excess emission is observed for about 1 year around the periastron passages only interrupted by a strong decline of a few weeks around $t = 0$ days. The light curves leading up to periastron are similar for each cycle; the brightness increases in all bands followed by a strong decline to a minimum. The recovery after the decline, however, differs from cycle-to-cycle. For example, in 1998, the recovery is slow and almost linear, while in 2003.5, on the same time scale, the light curve had quickly reached another maximum and was decreasing again.  
At apastron, when the influence of the companion star is minimal, the light curves also differ considerably from cycle-to-cycle. During the 2000 apastron there was a local minimum in $\eta$ Car's near-IR brightness, while 5.5~yrs later during the next apastron in 2006 the light curves showed a local maximum. 

Unfortunately, only the 1998.0 and the 2003.5 periastron passages and few mid-cycles are well-sampled in the near-IR. Nevertheless, together with the less frequent observations in other cycles, it is clear that the light curves vary from cycle to cycle making it difficult to fit one single model to the data. Most likely different processes, such as photoionization of the primary wind and a wind cavity, contribute in varying degrees.

\section{Discussion}
\label{discussion}

The $JHKL$ light curves from 1972 to 1998 show an almost linear increase in brightness, which can be explained with a simple model of dust dispersion as the Homunculus nebula expands  (e.g., \citealt{1971MNRAS.154..415D,1994A&A...283...89V}).
We find that prior to 1998 $\eta$ Car brightened at a rate of about $0.022$--$0.031$~mag~yr$^{-1}$ in $JHK$, similar to its brightening rate in the visual. \citet{1994A&A...283...89V} found that the star brightened almost linearly in $V$ between 1952--1992 at a rate of $0.025$~mag~yr$^{-1}$. In a simple model of an expanding Homunculus nebula, this near-neutral extinction implies large dust grains. 
The much slower brightening at $L$ of only $0.003$~mag~yr$^{-1}$ is presumably due to the reduced effects of extinction and a stronger contribution by thermal dust emission from the Homunculus \citep{2010MNRAS.402..145S}.  

The accelerated brightening in the UV to near-IR wavelengths after 1998 is inconsistent with this simple model (e.g., \citealt{1999AJ....118.1777D,1999A&A...346L..33S,2005AJ....129..900D}). \citet{2006JAD....12....3V} and \citet{2004MNRAS.352..447W} noted that the accelerated brightening may be related to the 1998 periastron passage.  Because $\eta$ Car is close to the Eddington limit, it is unlikely that the observations can be accounted for by an intrinsic brightening of the star. \citet{1999AJ....118.1777D} argued that the hot dust near the star (at a distance of a few hundreds of AU) must have decreased. Either existing dust was destroyed and/or the dust formation rate slowed. 
Plausible causes for the recent brightening involve changes in the stellar wind and its emergent radiation field, which require some change in the star itself. A decrease in mass-loss rate was proposed  \citep{2010ApJ...717L..22M}. 
\citet{2009ApJ...701L..59K,2010ApJ...717L..22M}, and \citet{2010ApJ...725.1528C} also inferred a drop in $\eta$ Car's mass-loss rate based on $\eta$ Car's X-ray light curve. 

Observations in the far-IR reinforce the idea of an altered stellar system. 
\citet{2013A&A...549A..67G} found that $\eta$ Car's 70--500~$\mu$m fluxes in 2010 are about a factor of 2 lower than expected from observations obtained 15 years earlier. They argue that the expansion of the Homunculus can only account for a 20\% drop in the far-IR radiation and propose dynamical changes in the circumstellar dust around $\eta$ Car.
However, \citet{2010MNRAS.401L..48G} found that the flux at 870~$\mu$m, a wavelength region dominated by free-free emission from the ionized wind, was a factor of 3 higher in 2007 compared to 1998, which can be primarily accounted for by the cyclical 5.5~yr variations. Radio monitoring by White (priv. comm.\footnote{See
http://www.astro.umd.edu/$\sim$white/images/eta\_time\_full.html and \citet{2005ASPC..332..126W}.}) shows that the radio flux at 8.6~GHz (3.5~cm) in 2006 was around 3~Jy, three times more than in 1998. However, the peak flux in cycle 1998--2003.5 was only at 2~Jy, supporting a discontinuity around 1998 in the radio as well.

Eta Car shows an intriguing color evolution in the near-IR that appears to be triggered close to periastron. 
The color $H$--$K$ progressed to the blue since the beginning of the near-IR observing record in the early 1970s. Figure \ref{color-color} shows that this blueward evolution in $H$--$K$ occurred discontinuously  within a short time span of only a few weeks close to the periastron passages. 
Before 1998, $J$--$H$ was basically constant around a value of  about 1~mag but is evolving to the blue since.  

The colors corrected for the long-term brightening trend (see Figure \ref{cycle-trendcorr} and its discussion in Section \ref{results}) are suggesting that we may be witnessing a temperature change in the extended tail of $\eta$ Car's wind from 3500~K to 6000~K as the optically thick near-IR emission region is moving closer to the star and/or the wind temperature is increasing. 
Support for this idea comes from \citet{2003A&A...410L..37V} and \citet{2007A&A...464...87W}. \citet{2003A&A...410L..37V} estimated from interferometric VLTI/VINCI observations in 2002 and 2003 that half of the K band flux comes from within 11~AU and estimated a minimum blackbody temperature of 2300~K concluding that they have spatially resolved the ionized stellar wind.  \citet{2007A&A...464...87W} found from  interferometric VLTI/AMBER observations in 2004/2005 an image diameter of the K-band continuum corresponding to about 5~AU, where temperatures of less than 8000~K are expected.

\subsection{Possible scenarios for the observed changes in $\eta$~Car}

It is highly improbable that the process of dust dissipation by the expanding Homunculus nebula is the sole cause for the above described observations. 
To explain the recent magnitude and color trends a second process needs to be invoked. This second process can be either related to bremsstrahlung emission or to hot dust, and is geometrically distinct from the dust clearance process occurring within the Homunculus nebula.
The dramatic change in $\eta$ Car's light curves in 1998 and the correlation of the discontinuous near-IR color evolution with periastron passages suggest that the close approach of the companion star is one of the main drivers for the observed changes. It appears conceivable that the close approach of the companion induces alterations in $\eta$ Car's atmosphere and its close environment. 
The long-term color changes support a change in the size/temperature of the optically thick wind,
although less warm dust can be considered a viable alternative. We will
evaluate two scenarios that may not be exclusive and may also not be the only conceivable ones.
For the discussion of the long-term evolution here, we do not consider the existence of optically thin free-free emission and its variability, which may drive the cyclic variability.

{\it \noindent Scenario I: Homunculus dust dissipation + bremsstrahlung emission.}

For following reasons we suggest that the preferred mechanism is a combination of thinning dust as the Homunculus nebula expands in combination with variability in  optically thick bremsstrahlung emission.
\begin{enumerate}
\item Photometric variability occurs simultaneously across $JHKL$. 

\item Color evidence for an increased temperature.

\item Eta Car's visual extinction is $A_{\textnormal{\scriptsize{V}}} \sim 6$--$7$~mag \citep{2006ApJ...642.1098H} and small  ($<$1--2~mag) at $L$. Thus the relative fluxes between $V$ ($\approx 4.5$~mag in 2013) and $L$ ($\approx -1.6$~mag in 2013) are consistent without invoking a large dust emission component in $L$.

\end{enumerate}

Dust production in a wind-wind collision region as observed in Wolf-Rayet (WR) stars does not work well for $\eta$ Car. Eta Car is more luminous than WR stars and even with an order of magnitude or two denser wind, its UV radiation will permeate through its wind. There are also a supposedly hot ($T_{\textnormal{\scriptsize{eff}}}  \sim 37\,000$~K) companion and colliding wind shocks as additional sources of ionizing photons and energy. 
It is thus unlikely that dust will form within 150~AU as the local thermal environment is too harsh for dust formation ($T_{\textnormal{\scriptsize{eff}}} \sim 1500$~K at 150~AU). Likely, the dust-free region around $\eta$ Car extends out to 200--300~AU, which is also inferred from high-angular resolution imaging, see \citet{2005A&A...435.1043C}. Considering that the significantly dense part of the wind-collision region is contained within 100~AU, a rapid (on a time scale of days to months) dust formation will likely not occur. 
Also, contrary to dust formation in WR stars \citep{2013MNRAS.429..494W}, the largest amplitudes of variability in the near-IR of $\eta$ Car are found in $J$ rather than in $L$-band. 

If we assume the near-IR emission is dominated by optically thick plasma, then the observed drop in $L$-band flux below the long-term trend could have been caused by a decreasing effective radius. Here ``effective radius'' means the radius at which the optical depth of the plasma changes from optically thick to thin.
A smaller effective radius could have resulted from a lower mass-loss rate, leading to an increased temperature at the effective radius. This is accounted for by the increased $J$ and $H$ emission above the long-term trend. The $K$ band around 2.2~$\mu$m may be the pivotal wavelength around which the bound-free and free-free emission component hinges: the $K$-band flux decreases because the emitting volume becomes smaller but this decrease is compensated for by the temperature increase. Thus the $K$ band flux is about constant (when corrected for the long-term brightening trend associated with the expanding Homunculus nebula).

We can further speculate on the physical causes for a possible change in mass-loss rate and the importance of the 
periastron passages.
A possible hypothesis for a physical cause is angular momentum transfer between the two stars at periastron. 
At the point of the closest approach angular momentum could be transferred onto the primary through tidal torques that spin up $\eta$ Car's atmosphere \citep{1997NewA....2..387D}. Incidentally,  $\eta$ Car's rotational velocity is most likely close to its critical velocity  now. 
\citet{2010ApJ...716L.223G} found in VLTI/VINCI and AMBER observations that $\eta$ Car's rotation speed is 80--90\% of its critical rotation speed.\footnote{However, \citet{2010ApJ...716L.223G} note that the primary wind could be sufficiently disturbed by the companion to mimick the effects of fast rotation.} 
 The angular momentum transfer scenario appears conceivable since $\eta$ Car's rotational velocity is likely close to the orbital velocity of the companion star, i.e., $v_{\textnormal{\scriptsize{rot}}}  \sim$ 0.8--0.9$\times v_{\textnormal{\scriptsize{crit}}} \sim$ 240--270~km~s$^{-1}$, whereas $v _{\textnormal{\scriptsize{orb}}} \sim$ 250--300~km~s$^{-1}$
 (with a companion mass of $\sim$ 40--50~$M_{\odot}$ and an orbit eccentricity of 0.8 or higher). 
Eta Car may have been approaching its critical velocity through an impulsive increase of $v_{\textnormal{\scriptsize{rot}}}$ at every periastron. The increase occurred also prior to 1998, but the rotational speed then may not have been close enough to its critical limit to induce the more dramatic changes observed afterward. This scenario could have been potentially enhanced, especially after 1998, if the star has expanded because of an internal instability \citep{1997NewA....2..387D}. However, the nature of this internal instability is not clear to the authors. A tidal spin-up of the star may be sufficient to trigger some kind of rotational instability and it may not be necessary to invoke an additional internal instability.

\citet{2005ASPC..332..169O} suggested that the rapid (near-critical) stellar rotation of $\eta$ Car induces an equatorial gravity darkening. 
While the radiative flux itself scales in proportion to the effective surface gravity, the mass flux scales directly with effective gravity, i.e., it is, contrary to intuitive expectations, at minimum near the equator and at maximum near the poles \citep{1996ApJ...472L.115O}.   
It may be that the companion star spins up $\eta$ Car incrementally on a regular basis (in addition to any spinning mechanism that may be present in LBVs). Its rotational velocity is further approaching its critical velocity, leading to lower effective gravity at the equator, and thus reducing the mass-loss rate there.\footnote{Note that \citet{2001RMxAC..10..199K} argued that angular momentum transfer from the companion briefly enhances $\eta$ Car's mass-loss rate during periastron. However, this effect would only occur briefly at the point of closest approach.} 
In view of this scenario, the $JHK$ excess emission during periastron passages can be explained by invoking a disruption of the primary wind, effectively mimicking a decrease in mass-loss rate (i.e., hotter zones of the optically thick wind are exposed), see \citet{2012ASPC..465..313M} and Section \ref{disc:excess}.

Soker et al.\ have discussed in a series of papers an accretion model for $\eta$ Car, which argues for the opposite effect that we just stated (see, e.g., \citealt{2005ApJ...635..540S,2008NewA...13..569K,2009NewA...14...11K}). In their scenario, the companion accretes mass from $\eta$ Car during periastron. In this case angular momentum from $\eta$ Car is transferred onto the companion, resulting in a slowing of its stellar rotation. However, the net change in angular momentum of $\eta$ Car depends on the mass fraction of the wind accreted onto the companion. The extended part of $\eta$ Car's wind carries only a negligible amount of the star's total mass, hence there may be a net gain in the angular momentum upon periastron passage.

{\it \noindent Scenario II: Homunculus dust dissipation + hot dust formation.}

The results in the near- to far-IR suggest that the inner dust envelope, enshrouding the central star, is opening up. A larger fraction of the stellar UV to near-IR radiation, which was previously absorbed within the inner dust envelope, is leaving the system, leading to less thermal dust emission at far-IR wavelengths and increased flux at wavelengths dominated by $\eta$ Car's stellar wind.

In principle, the  variations and post-1998 behavior could be explained with a scenario of dust removal at periastron and dust formation/recovery at apastron, but there are caveats related to 1.) the dust formation environment and 2.) the dust formation time scale. From the decreasing $K$ and $L$-band flux below the long-term trend in the last few years (though the effect in the light curve appears to be minute) we may infer that the amount of hot dust ($T_{\textnormal{\scriptsize{dust}}} \sim$1000-1500~K) must have become less since 1998, lowering the total extinction allowing more $JH$ flux to escape. As this is triggered at periastron, the host dust needs thus to be cleared at periastron, where a wind-collision dust production would be most effective. Prior to 1998, at post-periastron, the hot dust must have newly formed in the course of about one year as the near-IR fluxes levels are observed to revert to the pre-periastron state and this (newly formed) hot dust needs to be homogeneously distributed throughout the environment quickly after periastron passage.  Post-1998, the hot dust does not reappear or at least in much smaller quantities. This could be explained with a change in the wind-collision properties, e.g., different densities. If so, then we can assume that different densities occur in the primary's wind post-1998 and we arrive at the same conclusion as in Scenario I, namely that $\eta$ Car's mass-loss rate must have decreased.

In a slightly different scenario, we may assume that the hot dust is formed in the primary wind at $R_{\textnormal{\scriptsize{subl}}} \sim$150~AU, but then the question remains why the dust stopped forming post-1998.
Note that in Scenario II reasoning along the lines of dust in supergiant B[e] stars (gravity darkening, dense equatorial outflow with dust formation) is hard to uphold, because we need the hot dust to be spherically distributed or at least to be present in our line-of-sight.

\subsection{Excess emission around the periastron passages}
\label{disc:excess}

Eta Car's $JHKL$ light curves show excess emission correlated with the times of the periastron passages. 
\citet{2012ASPC..465..313M} proposed a model where the companion introduces a wind cavity in $\eta$ Car's dense wind. This cavity leads to increased radiation from the deeper layers of $\eta$ Car's wind close to periastron. They were able to replicate the light curves around periastron including the excess brightness and the miminum.

Alternatively, \citet{2008NewA...13..569K} and \citet{2010MNRAS.402..145S} proposed the formation of hot dust ($T_{\textnormal{\scriptsize{dust}}}  \sim 1500$~K) in the colliding winds to explain the increased near-IR flux around periastron. 
 \citet{2005MNRAS.357..895F} also showed that conditions in their model are favourable for grain formation.
However, it is difficult to sustain a large amount of dust at $T_{\textnormal{\scriptsize{dust}}}  \sim 1500$~K in the harsh environment around $\eta$ Car. \citet{2004MNRAS.352..447W} rejected an orbital cycle of grain growth and destruction since more dramatic variations in $L$ band than observed would be expected and  \citet{2005A&A...435.1043C}  found in near-IR high-angular  resolution (60~mas resolution) VLT/NACO images a dust-free area  with a radius of 230--350 AU around $\eta$ Car.
Other explanations for the excess emission at periastron include S Doradus-type variations \citep{1994MNRAS.270..364W,1999A&A...346L..33S}, enhanced free-free emission as newly ejected mass joins the material around $\eta$ Car \citep{2001MNRAS.322..741F}, and expansion of the primary due to tidal forces \citep{2006JAD....12....3V}.  
It is also possible to envisage the excess emission at periastron as a consequence of grain destruction. Increased UV radiation from deeper layers of $\eta$ Car's wind close to periastron may progressively heat large amounts of dust particles up to sublimation temperature of about 1500~K.

\section{Conclusion}
\label{conclusion}

We presented $JHKL$ photometry of  $\eta$ Car obtained since 2004
and compared the post-1998 data to the historical record from
1972­--1998.  We find that the near-IR light curves have considerably
altered their long-term behaviour around 1998 with respect to the general trend from
before 1998.
The change is most pronounced in $J$ and $H$, where excess emission above the
long-term trend is observed. The near-IR data  also show strong
episodic and long-term color variations. Sudden
near-IR color changes appear close to periastron passages. While the pre-1998 colors
were practically unchanged when corrected for the long-term brightening trends (which we associate with the expansion of the Homunculus nebula), the
post-1998 colors
evolve perceptibly toward the blue. The observed changes are
inconsistent with a gradual
dissipation of dust located in the Homunculus nebula as the sole cause and we
discuss
other processes that could explain the observations.

We argue that the changing near-IR emission is plausibly caused by
variability in optically thick bremsstrahlung emission. This is supported both by the discontinuity in the long-term trend since 1998 and the shorter term variations related to the
binary orbit.  Note that the existence of optically thin free-free emission components is also likely, which may contribute to the cyclic variation of colors, especially at the time of periastron passage.
We show by means of color-color analysis that the variability
may imply that a thermal component, presumably the stellar wind, has become
hotter and/or the boundary radius between optically thick and thin emission has become smaller. This suggestion is consistent with recent changes
in $\eta$
Car's UV, optical, and X-ray light curves and its optical stellar wind
features. These
observational findings all support an interpretation in terms of a
decrease in $\eta$ Car's
mass-loss rate resulting in more ionizing photons permeating its stellar
wind.

The periastron passages, especially the one in 1998, may be inferred to be
the driving force behind the observed quasi-periodic changes related
to the orbital period. The fine temporal sampling of the near-IR radiation strongly
suggests that periastron passages coincide with discontinuities in the long-term
trend and therefore related  to a persistent change in the system. This view is also supported by the
spectroscopic record. 
As a possible physical cause we propose that angular momentum transfer between the two stars at periastron may be altering $\eta$ Car's stellar rotation and then in turn its mass-loss rate. This may imply that $\eta$ Car is becoming more unstable as it nears its critical rotation limit, alluding to the possibility of another eruptive phase.

Evidence is increasing that the periastron passages play a key role in the evolution of the system. Further modelling of the interaction of the companion with material close to  $\eta$ Car and with its wind during periastron is required to disentangle the
various physical processes.
From an observational point, near-IR spectral monitoring is needed to
provide information
on the variability of the \ion{H}{I} and \ion{He}{I} emission lines and how they
influence the observed photometry. Unfortunately, no far-UV data was obtained since 2003
but MAMA observations of $\eta$ Car with {\it HST\/} within the next year
will hopefully determine if a drop in mass-loss rate occurred.
Observations at all wavelengths throughout the next periastron
in 2014.5 will be essential. The structure of $\eta$  Car is complex and
the interpretation
of the near-IR data is thus difficult. Nevertheless, important evidence
for the evolution
of this system can be obtained from the integrated near-IR photometry as
presented in this
paper.

\begin{acknowledgements} 
This paper is based on observations made at the South African Astronomical Observatory (SAAO). AM was co-funded under the Marie Curie Actions of the European Commission (FP7-COFUND). KI was supported by the Global Center of Excellence Program of Nagoya University from the Japan Society for the Promotion of Science and the Ministry 
    of Education, Culture, Sports, Science and Technology.  PAW and MWF acknowledge the receipt of a research grant from the South African National Research Foundation (NRF).
We thank Jose Groh, Mikio Kurita, Kris Davidson, and Stan Owocki for valuable discussions. 
\end{acknowledgements}


\begin{longtable}{llllll}
\caption{\label{table:phot} $JHKL$ photometry of $\eta$ Car with Mk~II and $JHK_{\textnormal{\scriptsize{s}}}$ photometry with SIRIUS at SAAO.}\\
\hline\hline
Date	&	MJD & J & H &  K & L \\    
(UT) & (days) & (mag) & (mag) & (mag) & (mag)  \\    
\hline
\endfirsthead
\caption{continued.}\\
\hline\hline
Date	&	MJD & J & H &  K & L \\    
(UT) & (days) & (mag) & (mag) & (mag) & (mag)  \\    
\hline
\endhead
\hline
\endfoot
\multicolumn{6}{c}{Mk~II photometry at the 0.75-m telescope\tablefootmark{a}}  \\
2004 Jan 27	&	53031.1	&	2.312	&	1.425	&	0.344	&	-1.686	\\
2004 Feb 25	&	53060.0	&	2.379	&	1.503	&	0.412	&	-1.655	\\
2004 Feb 27	&	53062.0	&	2.354	&	1.496	&	0.410	&	-1.676	\\
2004 Feb 28	&	53063.0	&	2.346	&	1.501	&	0.414	&	-1.676	\\
2004 Feb 29	&	53064.0	&	2.354	&	1.488	&	0.393	&	-1.697	\\
2004 Mar 01	&	53065.0	&	2.372	&	1.500	&	0.404	&	-1.653	\\
2004 Mar 01	&	53065.9	&	2.366	&	1.493	&	0.404	&	-1.638	\\
2004 Mar 02	&	53066.9	&	2.371	&	1.487	&	0.396	&	-1.669	\\
2004 Mar 04	&	53068.0	&	2.363	&	1.479	&	0.396	&	-1.690	\\
2004 Mar 06	&	53070.0	&	2.348	&	1.489	&	0.391	&	-1.660	\\
2004 Apr 20	&	53115.8	&	2.337	&	1.484	&	0.406	&	-1.673	\\
2004 Apr 21	&	53116.9	&	2.345	&	1.478	&	0.390	&	-1.656	\\
2004 Apr 24	&	53119.9	&	2.338	&	1.474	&	0.392	&	-1.654	\\
2004 Apr 25	&	53120.9	&	2.356	&	1.477	&	0.394	&	-1.648	\\
2004 May 13	&	53138.8	&	2.338	&	1.452	&	0.384	&	-1.636	\\
2004 May 16	&	53141.8	&	2.324	&	1.456	&	0.382	&	-1.680	\\
2004 Jun 15	&	53171.8	&	2.319	&	1.462	&	0.383	&	-1.664	\\
2004 Jun 23	&	53179.7	&	2.329	&	1.467	&	0.384	&	-1.686	\\
2004 Jun 28	&	53184.7	&	2.335	&	1.451	&	0.377	&	-1.644	\\
2004 Jul 07	&	53193.7	&	2.315	&	1.455	&	0.381	&	-1.651	\\
2004 Jul 11	&	53197.7	&	2.317	&	1.454	&	0.385	&	-1.665	\\
2004 Jul 16	&	53202.7	&	2.317	&	1.455	&	0.389	&	-1.691	\\
2004 Jul 19	&	53205.7	&	2.329	&	1.466	&	0.396	&	-1.674	\\
2004 Aug 17	&	53234.7	&	2.289	&	1.440	&	0.364	&	-1.679	\\
2004 Sep 03	&	53251.2	&	2.305	&	1.453	&	0.385	&	-1.676	\\
2004 Sep 07	&	53255.2	&	2.305	&	1.451	&	0.392	&	-1.671	\\
2004 Sep 29	&	53277.2	&	2.283	&	1.429	&	0.368	&	-1.663	\\
2004 Oct 12	&	53290.2	&	2.267	&	1.448	&	0.365	&	-1.697	\\
2004 Nov 04	&	53313.1	&	2.289	&	1.457	&	0.389	&	-1.696	\\
2004 Nov 18	&	53327.1	&	2.250	&	1.435	&	0.370	&	-1.725	\\
2004 Dec 09	&	53348.1	&	2.267	&	1.419	&	0.350	&	-1.700	\\
2005 Feb 02	&	53403.1	&	2.214	&	1.379	&	0.320	&	-1.728	\\
2005 Feb 06	&	53407.0	&	2.237	&	1.393	&	0.330	&	-1.701	\\
2005 Feb 11	&	53412.0	&	2.214	&	1.370	&	0.316	&	-1.726	\\
2005 Feb 15	&	53416.0	&	2.232	&	1.384	&	0.324	&	-1.700	\\
2005 Mar 01	&	53430.0	&	2.195	&	1.371	&	0.317	&	-1.737	\\
2005 Mar 08	&	53437.0	&	2.221	&	1.391	&	0.336	&	-1.723	\\
2005 Mar 12	&	53441.0	&	2.225	&	1.398	&	0.344	&	-1.701	\\
2005 Mar 23	&	53452.0	&	2.223	&	1.400	&	0.321	&	-1.717	\\
2005 Mar 28	&	53457.9	&	2.237	&	1.405	&	0.341	&	-1.712	\\
2005 Apr 09	&	53469.9	&	2.264	&	1.442	&	0.363	&	-1.712	\\
2005 Jul 06	&	53557.7	&	2.286	&	1.454	&	0.388	&	-1.663	\\
2005 May 15	&	53505.8	&	2.315	&	1.470	&	0.399	&	-1.650	\\
2005 May 16	&	53506.8	&	2.296	&	1.471	&	0.395	&	-1.667	\\
2005 Jul 08	&	53559.7	&	2.287	&	1.444	&	0.378	&	-1.662	\\
2005 Jul 11	&	53562.7	&	2.287	&	1.444	&	0.374	&	-1.660	\\
2006 Mar 06	&	53800.0	&	2.121	&	1.307	&	0.254	&	-1.740	\\
2006 Jun 03	&	53889.8	&	2.103	&	1.264	&	0.242	&	-1.703	\\
2006 Jun 05	&	53891.8	&	2.091	&	1.254	&	0.231	&	-1.705	\\
2006 Jun 15	&	53901.7	&	2.100	&	1.259	&	0.232	&	-1.723	\\
2006 Jul 17	&	53933.7	&	2.094	&	1.294	&	0.270	&	-1.705	\\
2007 Feb 08	&	54139.1	&	2.266	&	1.477	&	0.407	&	-1.633	\\
2007 Jul 05	&	54286.7	&	2.357	&	1.514	&	0.420	&	-1.612	\\
2008 Feb 15	&	54511.1	&	2.349	&	1.486	&	0.388	&	-1.656	\\
2008 Feb 19	&	54515.0	&	2.324	&	1.475	&	0.385	&	-1.668	\\
2008 Mar 14	&	54539.0	&	2.316	&	1.453	&	0.366	&	-1.694	\\
2008 Mar 19	&	54544.0	&	2.330	&	1.491	&	0.379	&	-1.707	\\
2008 Apr 23	&	54579.9	&	2.323	&	1.470	&	0.364	&	-1.690	\\
2008 Apr 24	&	54580.9	&	2.318	&	1.456	&	0.371	&	-1.691	\\
2008 Apr 26	&	54582.8	&	2.322	&	1.458	&	0.372	&	-1.685	\\
2008 Apr 27	&	54583.8	&	2.323	&	1.460	&	0.360	&	-1.694	\\
2008 Apr 28	&	54584.8	&	2.376	&	1.502	&	0.370	&	-1.707	\\
2008 Apr 29	&	54585.9	&	2.328	&	1.450	&	0.373	&	-1.678	\\
2008 May 27	&	54613.7	&	2.338	&	1.465	&	0.361	&	-1.695	\\
2008 Jun 24	&	54641.7	&	2.336	&	1.443	&	0.368	&	-1.659	\\
2008 Dec 25	&	54825.1	&	2.173	&	1.273	&	0.214	&	-1.770	\\
2008 Dec 26	&	54826.1	&	2.199	&	1.286	&	0.254	&	-1.699	\\
2008 Dec 31	&	54831.0	&	2.153	&	1.262	&	0.230	&	-1.711	\\
2009 Mar 03	&	54893.0	&	2.176	&	1.332	&	0.339	&	-1.634	\\
2009 Jun 27	&	55009.7	&	2.122	&	1.304	&	0.314	&	-1.635	\\
2010 Jan 30	&	55226.1	&	2.054	&	1.289	&	0.326	&	-1.607	\\
2010 Feb 05	&	55232.0	&	2.040	&	1.280	&	0.312	&	-1.640	\\
2010 Mar 24	&	55279.9	&	2.063	&	1.286	&	0.328	&	-1.595	\\
2010 May 20	&	55336.8	&	1.995	&	1.228	&	0.276	&	-1.668	\\
2010 Jun 06	&	55353.8	&	1.982	&	1.219	&	0.259	&	-1.667	\\
2011 Feb 24	&	55616.0	&	2.042	&	1.288	&	0.332	&	-1.607	\\
2011 Feb 27	&	55619.9	&	2.026	&	1.291	&	0.333	&	-1.638	\\
2011 Mar 02	&	55622.0	&	2.028	&	1.298	&	0.324	&	-1.655	\\
2011 Mar 20	&	55640.9	&	2.074	&	1.323	&	0.352	&	-1.574	\\
2011 Jun 20	&	55732.8	&	2.082	&	1.326	&	0.338	&	-1.580	\\
2012 Mar 06	&	55992.9	&	2.116	&	1.356	&	0.384	&	-1.572	\\
2012 Apr 18	&	56035.8	&	2.073	&	1.300	&	0.316	&	-1.684	\\
2012 Jul 12	&	56120.7	&	2.197	&	1.407	&	0.437	&	-1.525	\\
2012 Jul 17	&	56125.7	&	2.081	&	1.336	&	0.357	&	-1.575	\\
2013 Feb 01	&	56324.0	&	2.103	&	1.368	&	0.376	&	-1.617	\\
2013 Feb 06	&	56329.1	&	2.100	&	1.352	&	0.362	&	-1.628	\\
2013 Feb 07	&	56330.9	&	2.104	&	1.349	&	0.378	&	-1.632	\\
2013 Feb 21	&	56344.0	&	2.092	&	1.356	&	0.356	&	-1.575	\\
2013 Feb 22	&	56345.0	&	2.116	&	1.349	&	0.367	&	-1.622	\\
2013 Feb 23	&	56346.9	&	2.103	&	1.370	&	0.365	&	-1.628	\\
2013 Feb 26	&	56349.9	&	2.066	&	1.338	&	0.349	&	-1.661	\\
\hline\hline
\multicolumn{6}{c}{SIRIUS photometry at the 1.4-m IRSF telescope\tablefootmark{b}}  \\
2012 May 14	& 56061.3 & 2.012$\pm$0.002 & 1.381$\pm$0.002 & 0.446$\pm$0.004 & \\
2012 Jun 14	& 56092.7 & 1.986$\pm$0.008 & 1.363$\pm$0.008 & 0.440$\pm$0.013 & \\
2012 Nov 16	& 56247.1 & 1.904$\pm$0.009 & 1.289$\pm$0.009 & 0.352$\pm$0.015 & \\
2013 Feb 06	& 56329.9 & 2.026$\pm$0.009 & 1.410$\pm$0.009 & 0.420$\pm$0.015 & \\
2013 Feb 07	& 56330.9 & 2.005$\pm$0.009 & 1.406$\pm$0.009 & 0.491$\pm$0.014 & \\
2013 Feb 08	& 56331.9 & 2.008$\pm$0.009 & 1.411$\pm$0.009 & 0.411$\pm$0.014 & \\
2013 Feb 14	& 56337.1 & 2.010$\pm$0.008 & 1.409$\pm$0.009 & 0.461$\pm$0.014 & \\
2013 Feb 15	& 56338.9 & 2.024$\pm$0.009 & 1.415$\pm$0.008 & 0.438$\pm$0.014 & \\
2013 Feb 21	& 56344.0 & 2.018$\pm$0.008 & 1.406$\pm$0.008 & 0.456$\pm$0.013 & \\
2013 Feb 21	& 56344.9 & 1.974$\pm$0.009 & 1.394$\pm$0.008 & 0.419$\pm$0.014 & \\
2013 Feb 23	& 56346.9 & 1.937$\pm$0.008 & 1.385$\pm$0.008 & 0.448$\pm$0.013 & \\
2013 Feb 24	& 56347.9 & 1.990$\pm$0.008 & 1.388$\pm$0.008 & 0.447$\pm$0.013 & \\
2013 Feb 26	& 56349.9 & 1.978$\pm$0.008 & 1.388$\pm$0.008 & 0.420$\pm$0.013 & \\
2013 Mar 13	& 56364.8 & 1.969$\pm$0.038 & 1.376$\pm$0.011 & 0.413$\pm$0.031 & \\
2013 Mar 15	& 56366.0 & 1.975$\pm$0.028 & 1.335$\pm$0.040 & 0.430$\pm$0.013 & \\
2013 Mar 17	& 56368.0 & 1.968$\pm$0.062 & 1.368$\pm$0.038 & 0.463$\pm$0.071 & \\
2013 Mar 18	& 56369.0 & 2.099$\pm$0.041 & 1.458$\pm$0.015 & 0.433$\pm$0.032 & \\
2013 Mar 19	& 56370.1 & 2.118$\pm$0.036 & 1.474$\pm$0.007 & 0.345$\pm$0.023 & \\
2013 Mar 20	& 56371.0 & 2.090$\pm$0.051 & 1.455$\pm$0.033 & 0.434$\pm$0.040 & \\
2013 May 31 & 56443.7 &		2.045$\pm$0.031	& 1.415$\pm$0.001	&	0.442$\pm$0.009	&  \\
2013 Jun 09 & 56452.7 &		2.029$\pm$0.017	& 1.397$\pm$0.006	&	0.434$\pm$0.024	&  \\
2013 Jun 11 & 56454.8 &		2.045$\pm$0.049	& 1.397$\pm$0.003	&	0.426$\pm$0.035	& 	 \\
2013 Jun 16 & 56459.7 &		2.021$\pm$0.012	& 1.391$\pm$0.008	&	0.450$\pm$0.020	& 	 \\
\hline \hline
\multicolumn{6}{l}{Notes. \tablefootmark{(a)} Uncertainties are 0.03~mag in $JHK$ and 0.05~mag in $L$.} \\
\multicolumn{6}{l}{\phantom{Notes. }\tablefootmark{(b)} Magnitudes are in the SIRIUS $JHK_{\textnormal{\scriptsize{s}}}$ system, see Section \ref{obs}.} \\
\end{longtable}

\end{document}